\documentclass[prd,twocolumn,superscriptaddress]{revtex4}
\usepackage{graphicx}
\usepackage{amsmath}
\usepackage{adjustbox}
\usepackage{amsfonts}
\usepackage{amssymb}
\usepackage{color}
\usepackage{amscd}
\usepackage{bm}
\makeatletter
\renewcommand{\maketag@@@}[1]{\hbox{\m@th\normalsize\normalfont#1}}%
\makeatother
\begin{document}
\title{First order corrections to black hole thermodynamics: a simple approach enhanced}
\author{Yong Xiao}
\email{xiaoyong@hbu.edu.cn}
\author{Yue-Ying Liu}
\affiliation{Key Laboratory of High-precision Computation and Application of Quantum Field Theory of Hebei Province,
College of Physical Science and Technology, Hebei University, Baoding 071002, China}
\affiliation{Hebei Research Center of the Basic Discipline for Computational Physics, Baoding, 071002, China}

\begin{abstract}
 The inclusion of higher derivative terms in the gravitational action brings about corrections to the original forms of black hole solutions and thermodynamics. A simple and valuable approach has emerged over two decades ago, which states that the first order corrections to black hole thermodynamics, caused by higher derivative terms, can be achieved without explicitly solving the perturbed metric. However, a direct application of this approach may potentially yield incorrect answers as it comes to AdS black holes, which poses a challenge to the valuable approach. In this paper, we identify the subtlety associated with AdS black holes is that  certain higher derivative terms may alter the asymptotic structure of the spacetime. By taking this subtlety into account, we successfully enhance the original approach and resolve the mentioned problem. Furthermore, we make an interesting observation that the extremalization analysis that underlies this approach also plays a fundamental role in the perturbation theories of quantum mechanics and other disciplines.
 \end{abstract}

 \maketitle

\section{Introduction}

In the framework of effective field theory, all possible diffeomorphism invariant curvature terms can be incorporated into the gravitational action. These terms can arise naturally from quantum fluctuations at short distances or from the low energy limit of string theories \cite{Boulware:1985wk,Cardoso:2018ptl,Clifton:2011jh}. To be specific, the effective action can be expressed as
\begin{align}
I=\frac{1}{16\pi}\int_\mathcal{M} (R-2\Lambda+\alpha \mathcal{L}_{\text{hd}}), \label{hiLagr}
\end{align}
where $\mathcal{L}_{\text{hd}}$ represents the terms with more than two derivatives acting on the dynamical fields, and $\alpha$ is its coupling constant. Typically, by treating these higher derivative terms as perturbations around Einstein gravity, one can obtain perturbed black hole solutions and investigate their thermodynamic properties.

In a notable study, Gubser et al. \cite{Gubser:1998nz} examined the higher derivative term
\begin{align}
\mathcal{L}_{\text{hd}}\!=\!C^{hmnk}C_{pmnq}C_{h}^{\ rsp}C^{q}_{\ rsk}\!+\!\frac{1}{2}C^{hkmn}C_{pqmn}C_{h}^{\ rsp}C^{q}_{\ rsk},
\label{lhd4order}
\end{align}
which is generated from type IIB string theory. They calculated the first order corrections to the thermodynamics of black branes, with the motivation to explain the well-known $3/4$ discrepancy between the weak and strong coupling limits of the free energy of the dual conformal field theory. During their analysis, the authors made an intriguing observation that, at first order in the coupling, the free energy can be directly obtained by substituting the unperturbed metric into the Euclidean integral, although, in principle, such quantities should be calculated using the perturbed metric. This simple approach was subsequently confirmed for AdS black holes with flat, spherical, and hyperbolic event horizons in various dimensions of spacetime \cite{Caldarelli:1999ar,Landsteiner:1999gb}.

Recently, this approach has been rediscovered in the analysis of asymptotically flat black holes \cite{Reall:2019sah,Xiao:2022auy,Ma:2023qqj}. In particular, without explicitly solving the perturbed metric, Reall and Santos proposed a method to derive the higher derivative corrections to Kerr black hole thermodynamics at the order $\mathcal{O}(\alpha)$. The method works for a general angular momentum $J$, which includes the special case of extremal black hole. In contrast, the traditional approach requires to solve the perturbed black hole solution, but, due to the complexity of the rotating metric, normally the spacetime geometry and thermodynamics can only be determined through a perturbation expansion in $J/M^2$ \cite{Cano:2019ore}. Therefore, this approach should not be considered merely as a trick, but rather as an indispensable technique that provides new information about the black hole physics that cannot be accessed using the traditional approach. Since Kerr black hole is not only a realistic object in astronomy but also a sensitive probe of new physics \cite{Horowitz:2023xyl}, the significance of the approach becomes self-evident.

However, a problem is that a direct application of this approach may potentially yield incorrect answers as it comes to AdS black holes. This counter-intuitive fact can be easily noticed by examining several examples of higher derivative terms $\alpha \mathcal{L}_{\text{hd}}$ added to the Lagrangian. For instance, although it produces accurate outcomes for certain higher derivative terms like that in eq.\eqref{lhd4order}, it fails to do so for other terms, such as the well-known Gauss-Bonnet term, by comparisons with exact solutions \cite{Hu:2023gru}. As a consequence, for an arbitrarily given higher derivative term, one loses control and confidence with the validity of this approach.

In this paper, we identify the subtlety associated with AdS black holes, i.e., the asymptotic structure of the spacetime, characterized by the effective cosmological constant, could be altered by certain higher derivative terms. We demonstrate how to appropriately take this issue into account and then apply the extremization analysis that underlies  Refs.\cite{Gubser:1998nz,Caldarelli:1999ar,Landsteiner:1999gb,Reall:2019sah,Xiao:2022auy,Ma:2023qqj}. Consequently, precise results can be achieved for all higher derivative terms, which enhances the original framework. We shall provide several examples to verify our approach, and further showcase its applications in topics such as extended black hole thermodynamics and weak gravity conjecture (WGC). In addition, we reveal a fundamental theoretical framework common to the perturbation theories of black hole thermodynamics, quantum mechanics, and other disciplines.

\section{Explanation of the simple approach}
\label{sec:explanation}

We make an explanation of the simple approach in this section. Actually, the validity of the simple approach was initially observed in Refs.\cite{Gubser:1998nz,Caldarelli:1999ar,Landsteiner:1999gb} as an empirical fact without a good theoretical explanation. However, in recent years, such a validity has been attributed to the extremization of the original Euclidean action by the unperturbed metric \cite{Reall:2019sah,Ma:2023qqj,Xiao:2022auy}.

For clarity, let us begin with explaining this approach for the simplest case of a 4-dimensional Schwarzschild black hole (with vanishing $\Lambda$). The perturbed metric caused by the higher derivative term $\alpha \mathcal{L}_\text{hd}$ can be expressed as $g_{\mu \nu}(T)=\bar{g}_{\mu \nu}(T)+\mathcal{O}(\alpha)$, where $\bar{g}_{\mu \nu}(T)$ represents the original Schwarzschild solution with the temperature $T$. To obtain the black hole thermodynamics, we need to evaluate the on-shell Euclidean action $I_{\text{tot}}=I_{\text{EH}}+\alpha I_{\text{hd}}$ by substituting $g_{\mu \nu}(T)$ into it. As a solution of the original Einstein gravity, $\bar{g}_{\mu \nu}(T)$ extremizes the Einstein-Hilbert part: $I_{\text{EH}}=\frac{1}{16\pi}\int_\mathcal{M} R +\frac{1}{8\pi}\int_{\partial \mathcal{M}}(K-K_0)$. So the perturbations around $\bar{g}_{\mu \nu}(T)$ only produce corrections at $\mathcal{O}(\alpha^2)$. Therefore, for calculations of $I_{\text{EH}}$ at the order $\mathcal{O}(\alpha)$, it makes no difference to use the perturbed metric $g_{\mu \nu}(T)$ or the unperturbed metric $\bar{g}_{\mu \nu}(T)$. This means that we can directly borrow the known expression $I_{\text{EH}}(\bar{g}_{\mu\nu})=-\frac{\beta \bar{M}(T)}{2}$ of the unperturbed Schwarzschild black hole, where $\beta\equiv 1/T$ and $\bar{M}(T)=\frac{1}{8\pi T}$. Meanwhile, for the higher derivative part: $\alpha\, I_{\text{hd}} =\frac{\alpha}{16\pi}\int_\mathcal{M} \mathcal{L}_{\text{hd}}$, it is also sufficient to use the unperturbed metric to evaluate it, because the parameter $\alpha$ is already present in this term. The boundary term corresponding to $ \alpha \mathcal{L}_{\text{hd}}$ vanishes, when evaluated on asymptotically flat solutions \cite{Reall:2019sah}.

In total, the Euclidean action for the perturbed black hole with temperature $T$ is given by
\begin{align}
    I_{\text{tot}} \doteq -\frac{\beta \bar{M}(T)}{2}+\alpha\, I_{\text{hd}}(\bar{g}_{\mu\nu}).\label{earlyads}
\end{align}
 Throughout the paper, we use the symbol ``$\doteq$" to indicate that the equality is valid at $\mathcal{O}(\alpha)$. By attaching $ I_{\text{tot}}$ to the partition function, the expressions of other thermodynamic quantities, such as the entropy $S(T)$ and energy $M(T)$, can be easily derived using standard formulas. In such a derivation, the only difficulty is that one should make a clear distinction between the barred and unbarred quantities. For instance, $\bar{M}(T)$ represents the mass (with known forms) of the original unperturbed black hole, while $M(T)$ represents the mass (to be calculated) of the perturbed black hole. Due to the effects of higher derivative terms, generally the form of $M(T)$ should be different from the known form of $\bar{M}(T)$.

Obviously, this approach is applicable to other types of black holes. For instance, the Euclidean action of the perturbed Kerr black hole with temperature $T$ and angular velocity $\Omega$ can be expressed as
\begin{align}
    I_{\text{tot}} \doteq -\frac{\beta \bar{M}(T,\Omega)}{2}+\alpha\, I_{\text{hd}}(\bar{g}_{\mu\nu}),
\end{align}
where $\bar{M}(T,\Omega)=\frac{1}{4 \pi T(1+ \sqrt{1+(\frac{\Omega}{2 \pi T})^2})}$ is the mass of the unperturbed Kerr black hole.

Here we explain why the perturbed metric $g_{\mu\nu}$ and the unperturbed metric $\bar{g}_{\mu\nu}$ should share the same temperature $T$ and angular velocity $\Omega$. The metric $g_{\mu\nu}$ represents a small deviation from $\bar{g}_{\mu\nu}$. It is essential that this deviation is not arbitrary; indeed, $g_{\mu\nu}$ must respect the boundary condition of $\bar{g}_{\mu\nu}$ in order to ensure the equality $I_{\text{EH}}(g_{\mu\nu})\doteq I_{\text{EH}}(\bar{g}_{\mu\nu})$. This has been analyzed in detail in Ref.\cite{Reall:2019sah} for the gravitational Euclidean action. From a mathematical perspective, for a reasonable comparison between $\mathcal{F}[f_1(x)]$ and $\mathcal{F}[f_2(x)]$ in functional analysis, both functions $f_1(x)$ and $f_2(x)$ must obey the same boundary conditions. Since we are working within the grand-canonical ensemble, we must keep $T$ and $\Omega$ fixed as part of the boundary conditions. We may also choose another thermodynamic ensemble for analytical convenience, such as one based on the parameters $T$ and $J$ \cite{Reall:2019sah}. Appendix A includes more discussions to clarify the subtleties regarding the boundary conditions.

\section{Analysis of AdS black holes}

The subtlety associated with AdS black holes is that the presence of higher derivative terms may lead to modifications in the asymptotic structure of the spacetime, which is characterized by the effective cosmological constant $\Lambda_e$. Actually, only in the cases where $\alpha \mathcal{L}_{\text{hd}}$ maintains the relation $\Lambda_e=\Lambda$, the approach described above can be directly applied. Below we will provide a careful analysis for both cases with either $\Lambda_e=\Lambda$ or $\Lambda_e\neq \Lambda$.

Upon explicit calculation, the Euclidean action of AdS black holes is divergent, which can be suitably regularized by subtracting the contribution of the pure AdS background.  Specifically, the regularized integral can be defined as the form
\begin{align}
    \int_{\mathcal{M}}^{(\text{Reg.})}\equiv \beta \lim\limits_{r_c\rightarrow \infty} \left(\int_{\mathcal{V}_{r_c}}^{(\text{BH})}-\sqrt{\frac{g_{tt}^{\text{BH}}(r_c
)}{g_{tt}^{\text{AdS}}(r_c)}}\int_{\mathcal{V}_{r_c}}^{(\text{AdS})}\right),
\end{align}
where the factor of the second term ensures that the black hole and the background AdS spacetime have the same time periodicity at the cutoff radius $r_c$ ($\mathcal{V}_{r_c}$ represents the corresponding volume) \cite{hawkingpage,Dutta:2006vs}. Accordingly, the regularized Euclidean action is identified as
\begin{align}
    I_{\text{tot}}^R=I_{\text{EH}}^R+\alpha I_{\text{hd}}^R,
\end{align}
where
\begin{align}
    I_{\text{EH}}^R &\equiv \frac{1}{16\pi}\int_{\mathcal{M}}^{(\text{Reg.})}(R-2\Lambda), \label{einhil}\\
I_{\text{hd}}^R &\equiv \frac{1}{16\pi}\int_{\mathcal{M}}^{(\text{Reg.})}\mathcal{L}_{\text{hd}}.
\end{align}
This background subtraction method for regularization is a well-established approach for accurately determining the Euclidean action of AdS black holes. Originating from the seminal work \cite{hawkingpage}, it has been widely utilized in the literature for many years. Compared to the holographic renormalization method, its main advantage is that it does not involve the intricate boundary terms and counter-terms in the computation of the Euclidean action. However, as a subtraction method, it has the drawback of not revealing the potentially existing Casimir energy of the AdS background; nevertheless, this does not affect the thermodynamic behavior of interest. Additionally, it is worth noting that the method also applies to black holes with spherical, flat, and hyperbolic horizons \cite{Caldarelli:1999ar}, as well as to rotating black holes \cite{Gibbons:2004ai}.

Assume that the perturbed black hole solution around the Schwarzschild-AdS metric had been solved. Then the expression of the effective cosmological constant $\Lambda_e$ can be read from the asymptotic behavior $g_{tt}\sim 1+ \frac{r^{2}}{l_e^{2}} $ at infinity, where $l_e$ represents the effective AdS radius satisfying $\Lambda_e=-\frac{(D-1)(D-2)}{2l_e^2}$. Here we proceed in $D$ dimensions, considering its potential relevance in the context of AdS/CFT. Note the differences between $\Lambda$ and $\Lambda_e$: the former is the ``bare" cosmological constant present in the action, while the latter is the ``dressed" cosmological constant that is read from the spacetime metric.  In practice, $\Lambda_e$ may deviate from $\Lambda$ due to the presence of the higher derivative term $\alpha L_{\text{hd}}$.

In the cases where $\Lambda_e=\Lambda$, the simple approach can be applied as usual. The perturbed metric can be expressed as $g_{\mu \nu}(T,\Lambda)=\bar{g}_{\mu \nu}(T,\Lambda)+\mathcal{O}(\alpha)$, where $\bar{g}_{\mu \nu}(T,\Lambda)$ extremizes the Einstein--Hilbert part \eqref{einhil} of the action. Therefore, at first order in $\alpha$, $I_{\text{EH}}^{R}$ is given by the known form of the original Schwarzschild-AdS black hole:
\begin{align}
    I_{\text{EH}}^{R}(T,\Lambda) \doteq \frac{\beta\, \Sigma_{D-2} }{16\pi}\left(\frac{\bar{r}_h^{D-1}}{l^2} -\bar{r}_h^{D-3} \right),\label{iehlam}
\end{align}
where $\bar{r}_h$ represents the outer horizon radius of the unperturbed black hole. The expression of $\bar{r}(T)$ can be solved from the relation
\begin{align}
T=\frac{D-3}{4\pi \bar{r}_h}+\frac{(D-1)\bar{r}_h}{4 \pi  l^2}, \label{temrh}
\end{align}
and inserted into eq.\eqref{iehlam}. But it is unnecessary in practice, because we can use the chain rule of calculus to compute quantities such as $\frac{\partial I_{\text{EH}}^R }{\partial T}$. Still, $\alpha I_{\text{hd}}^R$ can be evaluated using the unperturbed metric as well. In total, we obtain the Euclidean action $I_{\text{tot}}^R$. Note that the formula \eqref{iehlam} corresponds to that of an AdS black hole with a spherical horizon. We refer the interested reader to Ref.\cite{Caldarelli:1999ar} for more general formulas with different topologies.

In the cases where $\Lambda_e\neq \Lambda$, we have to redo the above analysis, which is a missing piece in previous literature. Here the perturbed black hole solution possesses an effective cosmological constant $\Lambda_e$ that deviates from $\Lambda$. Thus, to comply with the asymptotic structure of the AdS spacetime, we should express $g_{\mu \nu}(T,\Lambda_e)$ around $\bar{g}_{\mu \nu}(T,\Lambda_e)$ rather than (\emph{by naive thinking}) $\bar{g}_{\mu \nu}(T,\Lambda)$, which gives $g_{\mu \nu}(T,\Lambda_e)=\bar{g}_{\mu \nu}(T,\Lambda_e)+\mathcal{O}(\alpha)$. Since $\bar{g}_{\mu \nu}(T,\Lambda_e)$ extremizes the part $\int^{(\text{Reg.})}_\mathcal{M} (R-2 \Lambda_e)$, rather than $\int^{(\text{Reg.})}_\mathcal{M} (R-2 \Lambda)$, we suggest the decomposition of the Lagrangian as $\mathcal{L}=\frac{1}{16\pi}\big [(R-2\Lambda_e)+\alpha \mathcal{L}_{\text{hd}}+2(\Lambda_e-\Lambda)\big ]$. Accordingly, the Euclidean action can be written as
\begin{align}
    I_{\text{tot}}^R \!=\! I_{\text{EH}}^R(T,\Lambda_e) \!+\! \alpha I_{\text{hd}}^R \!+\!  \frac{1}{16\pi}\int_\mathcal{M}^{(\text{Reg.})} 2(\Lambda_e \!-\! \Lambda). \label{modifyI}
\end{align}
As before, $I_{\text{EH}}^{R}(T,\Lambda_e)$ can be evaluated using eqs.\eqref{iehlam} and \eqref{temrh} with the parameters $\Lambda_e$ and $l_e$ replacing $\Lambda$ and $l$. And $\alpha I_{\text{hd}}^R$ can be calculated using the unperturbed metric.

Then, the crucial question is: can we obtain the value of $(\Lambda_e-\Lambda)$ without explicitly solving the perturbed black hole metric? Amazingly, the answer is yes. The equations of motion for dynamical fields can be derived by varying the action. Since we are only interested in the pure AdS solution in order to read off $\Lambda_e$, the possible matter sectors in the action can be temporarily neglected. Consider generic higher derivative terms given by $\mathcal{L}_{\text{hd}}=\sum_k \mathcal{L}_k$, where $k$ counts the number of derivatives acting on the metric. The trace of the gravitational field equation has the form \cite{Xiao:2022auy,Oliva:2010zd}
\begin{align}
  \frac{2\!-\!  D}{2} R +D\,\Lambda+\alpha\, \sum\limits_k\frac{k \!-\! D}{2}  \mathcal{L}_{k}+\alpha \,\nabla_\mu K^\mu=0.\label{prl}
   \end{align}
The total derivative term $\nabla_\mu K^{\mu}$ consists of $\nabla_\mu$ and $R_{\alpha \beta \mu \nu}$, and its concrete form could be rather complicated. However, substituting the pure AdS metric with $\Lambda_e$ into eq.\eqref{prl}, this term vanishes because of $\nabla_\rho R_{\alpha \beta \mu \nu}=0$ for a pure AdS metric. As a result, the most troublesome term in the field equation poses no obstacle, and eq.\eqref{prl} readily reduces to $-D\,\Lambda_e+D\,\Lambda+\alpha \sum_k \frac{k-D}{2} \mathcal{L}_{k}^{ (AdS_{\Lambda_e} )}=0$, which is a non-perturbative equality and of great interest itself. As for the calculations at $\mathcal{O}(\alpha)$, it is irrelevant whether to use the perturbed AdS metric or the original AdS metric to evaluate the third term. Hence, we arrive at
\begin{align}
    \Lambda_e - \Lambda \doteq \alpha \sum_k \frac{k-D}{2 D} \,\mathcal{L}_{\text{k}}^{(\text{AdS})}. \label{lambdaRelation}
\end{align}
Combining it with eq.\eqref{modifyI}, we finally get the complete expression for $I_{\text{tot}}^R$.

\section{Gracefully dealing with AdS black holes}\label{secExample}

Based on the above analysis, when confronted with an arbitrary higher derivative term $\alpha \mathcal{L}_{\text{hd}}$, the key step is to judge whether it would alter the effective cosmological constant or not. It is remarkable that our formula \eqref{lambdaRelation} can be easily utilized to fulfill this purpose.

Consequently, it becomes a two-step procedure for dealing with AdS black holes. First, one substitutes a pure AdS metric into the right hand side (RHS) of eq.\eqref{lambdaRelation} and assesses whether it vanishes. If it does, one can then directly apply the original approach described around eq.\eqref{iehlam}. But if it doesn't, one should be careful with the subtlety about the asymptotic structure of the spacetime and turn to our analysis around eq.\eqref{modifyI}.

Let us examine several specific higher derivative terms as examples. The first example is that in eq.\eqref{lhd4order}. In this case, the RHS of eq.\eqref{lambdaRelation} vanishes, which actually holds for any higher derivative terms constructed from Weyl tensor. This elucidates why the original approach was nicely observed and verified in the early literature \cite{Gubser:1998nz,Caldarelli:1999ar,Landsteiner:1999gb}. Another example is the Gauss-Bonnet term.  The RHS of eq.\eqref{lambdaRelation} solely vanishes in $D=4$ dimensions but does not in higher dimensions. This explains why a refined approach should be constructed for these cases \cite{Hu:2023gru}. Below we present two additional examples with $\Lambda_e\neq \Lambda$.

\subsection{Example I: corrections from the $R^2$ term}
We consider the first order corrections to the thermodynamics of a 5-dimensional Schwarzschild-AdS black hole caused by the higher derivative term
\begin{align}
   \mathcal{L}_{\text{hd}} = R^{\mu\nu \rho \sigma }R_{\mu\nu \rho \sigma }.
\end{align}
For this example, we obtain $\Lambda_e-\Lambda =-\alpha\frac{\Lambda_e^2}{9}$ from eq.\eqref{lambdaRelation}. According to our approach, the integral $ \int_{\mathcal{M}}^{(\text{Reg.})}(R-2\Lambda_e)$ is given by the known form $2 \pi^2\beta \left(-\frac{1}{6} \Lambda_e \bar{r}_h^4-\bar{r}_h^2\right)$ (see eq.\eqref{iehlam}). Meanwhile, the integrals $ \alpha \int_\mathcal{M}^{(\text{Reg.})}\mathcal{L}_{\text{hd}}$ and $\int_\mathcal{M}^{(\text{Reg.})}2(\Lambda_e-\Lambda)=-\alpha\int_\mathcal{M}^{(\text{Reg.})} \frac{2 \Lambda_e^2}{9}$ can be evaluated using the unperturbed metric, which gives
\begin{align}
    \alpha \int_\mathcal{M}^{(\text{Reg.})}\mathcal{L}_{\text{hd}}\!\doteq \!\beta\frac{\pi ^2 \alpha}{18} \left(13 \Lambda_e^2 \bar{r}_h^4-246 \Lambda_e \bar{r}_h^2+648\right), \\
    \int_\mathcal{M}^{(\text{Reg.})}2(\Lambda_e-\Lambda)\doteq \beta\frac{\pi ^2 \alpha}{18} \left(\Lambda_e^2\bar{r}_h^4+6 \Lambda_e \bar{r}_h^2\right). \label{lamelam}
\end{align}
Clearly, eq.\eqref{lamelam} is nonzero and should not be ignored.

Summing them together, we obtain the expression of $I^R_{\text{tot}}$, which in turn gives the free energy as $F= -I^R_{\text{tot}}/\beta$. Then we can derive the expressions for the entropy and energy as functions of $\bar{r}_h$ (or equivalently $T$) from standard formulas. Moreover, we can also derive the conjugate quantities associated with the cosmological constant $\Lambda$ and the coupling $\alpha$, if treating $\Lambda$ and $\alpha$ as independent thermodynamic variables, as done in extended black hole thermodynamics \cite{Kastor:2009wy,Dutta:2022wbh,Ahmed:2023snm,Xiao:2023lap}.

To avoid diverging from our main thread, the derived expressions for the thermodynamic quantities are given in Appendix B. And our results indeed align with those in Ref.\cite{Dutta:2022wbh}, which were obtained from the traditional method based on the perturbed metric. The detailed calculations mentioned in this and the next examples have been included in a Notebook, which can be downloaded from the auxiliary file of the arXiv version of the paper.

\subsection{Example II: four-derivative corrections to the thermodynamics of a charged black hole}

The approach can be generalized to the theories where gravity is coupled with matter fields. Below we explain how to derive the first order corrections to the thermodynamics of a Reissner-Nordstr{\"o}m-AdS (RN-AdS) black hole in $D=5$ dimensions. We consider the effective Lagrangian
\begin{align}
    \mathcal{L} = \frac{1}{16\pi}(R - 2\Lambda - \frac{1}{4}F^2 + \alpha\,\mathcal{L}_4),
\end{align}
where the four-derivative term $\mathcal{L}_4$ is given by
\begin{align}
  \mathcal{L}_4  \!= \! c_1 R_{\mu\nu\rho\sigma}R^{\mu\nu\rho\sigma}  \!+ \! c_2 R_{\mu\nu\rho\sigma} F^{\mu\nu} F^{\rho\sigma}  \!+ \! c_3 (F^2)^2  \!+ \! c_4 F^4,
\end{align}
with $F_{\mu\nu} \equiv \nabla_\mu A_{\nu} - \nabla_\nu A_{\mu}$, $F^2 \equiv F_{\mu\nu}F^{\mu\nu}$, and $F^4 \equiv F_{\mu\nu}F^{\nu}_{\ \alpha}F^{\alpha}_{\ \beta}F^{\beta\mu}$.

For this example, we deduce $\Lambda_e - \Lambda = -\alpha\, c_1 \frac{\Lambda_e^2}{9}$ from eq.\eqref{lambdaRelation}. According to our approach, we decompose the Euclidean action into three parts. The first part $\int_{\mathcal{M}}^{(\text{Reg.})}(R-2\Lambda_e-\frac{1}{4}F^2)$ is extremized by the RN-AdS solution with the metric $\bar{g}_{\mu\nu}=-\bar{f}(r)dt^2+1/\bar{f}(r)\,dr^2+r^2 d\Omega_3$ and the vector field $\bar{A}_{\mu}=\left(-\frac{\sqrt{3} q}{2 r^2},0,0,0,0\right)$, where $\bar{f}(r)=1-\frac{\mu}{r^2}+\frac{q^2}{4 r^4}-\frac{\Lambda_e r^2}{6}$. The extremalization analysis tells that it can be directly evaluated using the known results of the original RN-AdS black hole parameterized by the temperature $T$, electrostatic potential $\Phi$, and asymptotic structure characterized by $\Lambda_e$. Meanwhile, we can calculate the second and third parts, i.e., $\alpha \int_\mathcal{M}^{(\text{Reg.})}\mathcal{L}_{4}$ and $\int_\mathcal{M}^{(\text{Reg.})}2(\Lambda_e-\Lambda)=-\alpha \int_\mathcal{M}^{(\text{Reg.})} \frac{2 c_1 \Lambda_e^2}{9}$ using the unperturbed metric. In total, we get the free energy depending on $T$ and $\Phi$. However, it is much more convenient to change to the ensemble depending on $T$ and $q$ by Legendre transformation. In the end, the free energy $F(T,q)$ is obtained as
\begin{align}
\begin{split}
   & F \doteq \frac{5 \pi  q^2}{32 \bar{r}_h^2} \!+\! \frac{1}{48} \pi  \Lambda_e \bar{r}_h^4 \!+\! \frac{\pi  \bar{r}_h^2}{8}  \!+\! \alpha \frac{\pi }{2304 \bar{r}_h^8}\left[ c_1(\!-\! 387 q^4  \right.\\
    & \left. \!-\!  192 \Lambda_e q^2 \bar{r}_h^6  \!+\!  1728 q^2 \bar{r}_h^4  \!-\! 112 \Lambda_e^2 \bar{r}_h^{12}\!+\! 1920 \Lambda_e\bar{r}_h^{10}\right. \\
    &\left.\!-\! 5184 \bar{r}_h^8) \!+\! 216 q^2 \left(8 \bar{r}_h^4 \!-\! 3 q^2\right) c_2 \!-\! 648 q^4 (2 c_3 \!+\! c_4)\right],
\end{split}
\end{align}
where the expression of $\bar{r}_h$ can be determined from
\begin{align}
   T=\frac{1}{2 \pi  \bar{r}_h}-\frac{\Lambda_e \bar{r}_h}{6 \pi }-\frac{q^2}{8 \pi  \bar{r}_h^5}.
\end{align}
Then, utilizing the relation $dF=-SdT+\Phi dQ +\Theta_e d\Lambda_e+U_\alpha d\alpha$, we can derive the expressions of $S$, $\Phi$, $\Theta_e$ and $U_\alpha$, as well as $M=F+TS$. Here, $Q\equiv \frac{ \sqrt{3} \pi }{8}q$ is the conserved electric charge, and $\Theta_e $ and $U_\alpha$ are respectively the conjugate quantities associated with $\Lambda_e$ and $\alpha$ in the framework of extended black hole thermodynamics. We have conducted detailed calculations for all thermodynamic quantities using both the above approach and the traditional method that relies on the perturbed metric. The results agree remarkably well.

Now we pause to discuss the applications of our approach on WGC. In this context, higher derivative terms $\alpha L_{\text{hd}}$ are introduced as manifestations of quantum gravitational effects. The researchers switch on the coupling $\alpha$ for an extremal charged black hole to examine the change in entropy and energy. Typically, a decrease in energy $M_{\text{ext}}$ is anticipated, which realizes a state with $Q/M>1$ favored by the WGC \cite{Harlow:2022ich,Cremonini:2019wdk,Noumi:2022ybv}. A pivotal formula relevant to this topic is expressed as
\begin{align}
  \left(\frac{\partial M_{\text{ext}}(Q,\alpha)}{\partial \alpha}\right)_Q \!=\! -\! \lim\limits_{M\rightarrow M_{\text{ext}}}T\left(\frac{\partial S(M,Q,\alpha)}{\partial \alpha}\right)_{M,Q}, \label{dmdsrelation}
\end{align}
which was initially proposed in Ref.\cite{Goon:2019faz}. The derivations of this and other related formulas are somewhat involved in the literature, but we find the formula can be easily observed from the extended first law
\begin{align}
    dM=TdS+\Phi dQ +\Theta_e d\Lambda_e+U_\alpha d\alpha. \label{extendedrnads}
\end{align}
Obviously, one can observe from eq.\eqref{extendedrnads} that $dM_{\text{ext}}=U_\alpha^{\text{ext}}\, d\alpha$ by fixing $Q$ and $\Lambda_e$ and using extremal temperature $T=0$, which leads to $\left(\frac{\partial M_{\text{ext}}(Q,\alpha)}{\partial \alpha}\right)_Q=\lim\limits_{T\rightarrow 0}U_\alpha$. Meanwhile, we get $T d S=-U_{\alpha}\, d\alpha$ with fixed $M$, $Q$ and $\Lambda_e$, so the RHS of eq.\eqref{dmdsrelation} also reduces to $\lim\limits_{T\rightarrow 0}U_\alpha$. This observation effortlessly leads to the formula \eqref{dmdsrelation}. Following the same logic, we can further observe other relations, such as $\left(\frac{\partial M_{\text{ext}}(Q,\alpha)}{\partial \alpha}\right)_Q=-\Phi\left(\frac{\partial Q_{\text{ext}}(Q,\alpha)}{\partial \alpha}\right)_M$, which was proposed in Ref.\cite{Cremonini:2019wdk}. All these quantities are simply equal to $\lim\limits_{T\rightarrow 0}U_\alpha$, where
\begin{align}
    U_{\alpha}=\left(\frac{\partial{F}}{\partial \alpha}\right)_{T,Q,\Lambda_e}. \label{mulame}
\end{align}

In the study of WGC, one is concerned about the slight deviations from Einstein gravity, so eq.\eqref{dmdsrelation} is usually evaluated around $\alpha=0$, indicating that the first order calculation is adequate for the task.  In this regard, our approach is particularly valuable in this area of research. For any higher derivative terms at hand, our approach rapidly yields the free energy $F(T,Q,\Lambda_e)$. Then one can deduce the expressions of $\lim\limits_{T\rightarrow 0}U_\alpha$ from eq.\eqref{mulame}, which reveals whether the higher derivative terms lead to an increase or decrease in the mass of the extremal black hole.

Besides, there are certain intricacies in investigating WGC for AdS black holes. The thermodynamic quantities with the coupling parameter $\alpha$ switched on and off should be compared under the same boundary conditions.
 In cases where $\Lambda_e=\Lambda$, the comparison goes smoothly. But, in cases where $\Lambda_e\neq \Lambda$, the analysis becomes involved and there exists other prescriptions for examining WGC. Specifically, the prescription of Ref.\cite{Cremonini:2019wdk} involves re-scaling the black hole solution by defining a new time coordinate $\tilde{t}= \lambda(\alpha)\, t$, where the re-scaling factor is chosen to satisfy the requirement $\Lambda_e=\lambda(\alpha)^2\,\Lambda$ that drags $\Lambda_e$ back to $\Lambda$. In this scenario, we can recover their outcome by using the re-scaled quantities, such as temperature $\tilde{T} = T / \lambda(\alpha)$ and free energy $\tilde{F} = F / \lambda(\alpha)$. Correspondingly, the relevant quantity for analyzing WGC is $\tilde{U}_{\alpha}=\left(\frac{\partial{\tilde{F}}}{\partial \alpha}\right)_{\tilde{T},Q,\Lambda}$.

\section{A universal theoretical structure}

We have now established that the first order corrections to black hole thermodynamics caused by any higher derivative terms can be derived using the \emph{unperturbed} metric, even for AdS black holes. As we explained in Sec.\ref{sec:explanation}, the validity of the methodology relies on the extremum property in functional calculus. Although the validity appears to be a trivial fact in hindsight, when initially observed in Refs.\cite{Gubser:1998nz,Caldarelli:1999ar,Landsteiner:1999gb}, the method was not well understood and often regarded as surprising and miraculous. It is understandable that the complexity of specific gravitational problems could easily obscure the identification of such a simple theoretical structure.

Below we make an intriguing observation that the situation also occurs in the perturbation theories of quantum mechanics and other disciplines. Though the methodology of extremalization analysis plays a fundamental role in these kinds of perturbation theory, it could easily get unnoticed due to the complexities of the concrete physical problems.

We notice that there exists a similar assertion in quantum mechanics, which states that the first order correction to the energy can be achieved from the \emph{unperturbed} quantum state \cite{gsbook}. The derivation is as below. Consider a quantum mechanical system with perturbed Hamiltonian $\hat{H}=\hat{H}_0 +\alpha H^{\prime}$. We start from a (nondegenerate) unperturbed state $|\psi_n^{(0)}\rangle$ that satisfies $\hat{H}_0 |\psi_n^{(0)}\rangle=E_n^{(0)} |\psi_n^{(0)}\rangle$. The question is to evaluate the energy in the perturbed state $|\psi_n\rangle \doteq |\psi_n^{(0)}\rangle+\alpha |\psi_n^{(1)}\rangle$:
\begin{align}
E_n=\langle\psi_n|\hat{H}|\psi_n\rangle=\langle\psi_n|\hat{H_0}|\psi_n\rangle+\alpha \langle\psi_n|\hat{H}^{\prime}|\psi_n\rangle. \label{expectationvalue}
\end{align}
For calculations at $\mathcal{O}(\alpha)$, the second term can be evaluated using the unperturbed state $|\psi_n^{(0)}\rangle$, because it has already included the parameter $\alpha$. As for the first term in eq.\eqref{expectationvalue}, in principle we should use the perturbed state to evaluate it, but it proves to be unnecessary. The difference between $\langle\psi_n|\hat{H_0}|\psi_n\rangle$ and $\langle\psi_n^{(0)}|\hat{H_0}|\psi_n^{(0)}\rangle$ is equal to $\alpha E_n^{(0)} (\langle \psi_n^{(1)}| \psi_n^{(0)} \rangle+\langle  \psi_n^{(0)} | \psi_n^{(1)} \rangle)$, which vanishes due to the normalization condition of $|\psi_n\rangle$ and $|\psi_n^{(0)}\rangle$.

Inspired by the insights from our analysis of black hole thermodynamics, we believe that an extremalization analysis should be hidden behind the above assertion in quantum mechanics. In essence, the unperturbed state $|\psi_n^{(0)}\rangle$ extremizes the functional $\langle\psi|\hat{H_0}|\psi\rangle$ under the implicit constraint $\langle\psi|\psi\rangle=1$. This is the fundamental reason why the difference between $\langle\psi_n|\hat{H_0}|\psi_n\rangle$ and $\langle\psi_n^{(0)}|\hat{H_0}|\psi_n^{(0)}\rangle$ vanishes at the first order in $\alpha$. In a deeper sense, the extremization of $\langle\psi|\hat{H_0}|\psi\rangle- \lambda \langle\psi|\psi\rangle$ is equivalent to solving the unperturbed Schr\"{o}dinger equation, where $\lambda$ is introduced as a Lagrange multiplier.

We have further examined a conventional thermodynamic system, which is a simplified model with two energy levels $E_i=E_i^{(0)}+\alpha E_i^{(1)}$ $(i=1,2)$ and the distribution function $\rho(E_i)\propto e^{-\beta E_i}$. By a parallel thinking, we surely find an approach to write down the free energy at $\mathcal{O}(\alpha)$, which only involves the unperturbed distribution function $\rho(E_i^{(0)})\propto e^{-\beta E_i^{(0)}}$. In this case, the extremization problem corresponds to solving the thermodynamic equation $\frac{\partial S(\bar{E})}{\partial \bar{E}}=\beta $.

Thus, we have recognized the existence of a universal theoretical structure that underlies various kinds of perturbation problems in different disciplines. The presence of this structure guarantees that there exists a convenient tool for extracting the first order corrections to the physical quantities without really solving the perturbed system. Moreover, it also means that the $n$-th order corrections to the physical quantities can be obtained from the $(n\!-\!1)$-th order solutions. Identifying similar theoretical structures in other types of perturbation systems, even beyond the realm of physics, would be highly fascinating.

\section{Concluding remarks}

In summary, we have investigated the first order corrections to black hole thermodynamics resulting from the inclusion of higher derivative terms. The subtlety associated with AdS black holes is that the higher derivative terms may modify the asymptotic structure of the spacetime, which poses a significant challenge to the original approach of Refs.\cite{Gubser:1998nz,Reall:2019sah}. In this paper, we have proposed a two-step procedure to deal with AdS black holes, and the crucial step is to utilize our formula \eqref{lambdaRelation} to assess whether the effective cosmological constant $\Lambda_e$ would deviate from $\Lambda$. Our approach is applicable to all higher derivative terms, thereby unlocking the full potential of the approach \cite{Gubser:1998nz,Reall:2019sah} originated from two decades ago.

As the feature of the approach, it is unnecessary to solve the perturbed metric. Therefore, it becomes an indispensable technique in situations where the perturbed metric is difficult or even impossible to obtain, especially for Kerr and Kerr-AdS black holes. In these cases, such approach provides almost the only means of gaining valuable insights into certain thermodynamic behaviors of these black holes. Since black hole thermodynamics plays an essential role in the understanding of quantum gravity, the approach has broad applications on topics such as WGC, extended black hole thermodynamics, as well as the precision tests of the AdS/CFT correspondence \cite{Ma:2024ynp}.

It is worth noting that, slightly later than our work, another group \cite{Hu:2023gru} independently recognized the same problem concerning AdS black holes, and developed a different
strategy to improve this approach. Their method entails complicated field redefinitions and specifically deals with $4$-derivative terms. In contrast, our approach is more straightforward and applicable to all $n$-derivative terms with $n \geq 4$.

\ \
\section*{Acknowledgments}
YX is grateful to Harvey S. Reall, Jorge E. Santos, Alejandro Ruip\'{e}rez, Liang Ma and Hong L\"{u} for extensive discussions. We also thank the anonymous referee for the useful suggestions. This work was partially supported by the Hebei Natural Science Foundation (NSF) with Grant No. A2024201012, by NSF of China with Grant No. 12475048.

 \section*{Appendix A}
This appendix further clarifies the subtleties of the boundary conditions associated with this simple approach.

In the footnote $6$ of Ref.\cite{Reall:2019sah}, Reall and Santos have made an explanation of the subtleties related to AdS black hole. Specifically, in order to get the correct results of the Euclidean action, it is important to fix the boundary conditions of the perturbed metric to match those of the unperturbed metric. We elaborate on this issue below.

It is known that we can adjust the asymptotic behavior of the metric through time re-scaling, so we consider four metrics: (a) the perturbed black hole metric, which solves the field equations of the Lagrangian with higher derivative terms and has the standard form $g_{tt}\sim 1-\frac{\mu}{r^{D-3}}+\frac{r^2}{l_e^2}+\mathcal{O}(\alpha) $ possessing a corrected AdS radius $l_e$; (b) the re-scaled perturbed black hole metric with the form  $g_{tt}\sim \frac{1}{{\lambda}^2}-\frac{\mu/{\lambda}^2}{r^{D-3}}+\frac{r^2}{l^2}+ \mathcal{O}(\alpha) $, which restores the original AdS radius $l$ by introducing a re-scaling factor $\lambda$ that satisfies $\l=\lambda l_e$; (c) the unperturbed black hole metric with the form $g_{tt}\sim 1-\frac{\bar{\mu}}{r^{D-3}}+\frac{r^2}{l^2}$, which solves the original two-derivative Lagrangian and has the AdS radius $l$; and (d) the re-scaled unperturbed black hole metric with a form $g_{tt}\sim \lambda^2-\frac{\lambda^2 \bar{\mu}}{r^{D-3}}+\frac{r^2}{l_e^2}$, which has a corrected AdS radius $l_e$ through an appropriate re-scaling of the above metric.

Here are the key points: when the boundary conditions at $r\rightarrow \infty$ are the same, the simple method of \cite{Gubser:1998nz,Caldarelli:1999ar,Landsteiner:1999gb,Reall:2019sah} can be effectively applied. Concretely speaking, to obtain the first order Euclidean action for the black hole metric (a), one can use metric (d) along with the method from \cite{Gubser:1998nz,Caldarelli:1999ar,Landsteiner:1999gb,Reall:2019sah}. The derived results will align with those obtained by directly inserting metric (a) into the action, since the two metrics (a) and (d) share the same boundary conditions. Similarly, to obtain the Euclidean action for black hole metric (b), one can use metric (c) along with the simple method. These statements can be readily verified through concrete examples.

 The metrics (a) and (b) are linked by a time re-scaling, thus both can be viewed as solutions to the corrected field equations, selected by imposing specific boundary conditions. The usage of the two metrics depends on the motivation behind choosing them. For comparing the properties of a perturbed black hole with an unperturbed one, the metric (b) is more suitable, and its first order Euclidean action can be smoothly derived from metric (c). However, often our task is to obtain the thermodynamics of a black hole described by metric (a). For instance, the well-known Gauss-Bonnet metric is of type (a). It would be a challenge to use the method of Refs.\cite{Gubser:1998nz,Caldarelli:1999ar,Landsteiner:1999gb,Reall:2019sah} to obtain its Euclidean action without knowing the exact form of metric (a). In such cases, one needs to acquire the form of the metric (d) by re-scaling (c). But, unfortunately, the time re-scaling factor is determined by the concrete form of $\Lambda_e$ which is unknown without solving (a), as pointed out in \cite{Hu:2023gru}. Transparently, this obstacle has been overcome by our pivotal formula \eqref{lambdaRelation}.

As repeatedly emphasized in our main text, it should be a two-step procedure to calculate the Euclidean action of the metric (a). The crucial step is to judge whether $\Lambda_e$ deviates from $\Lambda$ using our formula \eqref{lambdaRelation}. If the situation is $\Lambda_e\neq \Lambda$, one should be cautious with the boundary conditions.  Depending on personal preference, one can adopt either our approach around eq.\eqref{modifyI} or the time re-scaling approach described above in this appendix. The advantage of using eq.\eqref{modifyI} is that it is purely a mathematical manipulation, avoiding touching the time coordinate. In contrast, using the time re-scaling approach, one has to keep a clear mind in the distinction among the original temperature and the re-scaled temperature, and so on.

 \section*{Appendix B}
This appendix explains how the thermodynamic quantities of the perturbed black hole can be derived from the Euclidean action $I_{tot}$. As mentioned, the results can be easily reproduced using our Notebook.

We continue with our analysis of the example I in Sec.\ref{secExample}. Starting from the free energy $F= -I^R_{tot}/\beta$, we can derive the entropy as
\begin{align}\label{B1}
    S=-\frac{\partial F}{\partial T}\doteq \bar{S}(T,\Lambda_e) \left( 1+ \frac{\alpha \left(60 \Lambda_e-7 \Lambda_e^2 \bar{r}_h^2\right)}{3 \left(\Lambda_e \bar{r}_h^2+3\right)} \right), \tag{B.1}
\end{align}
and the black hole energy as
\begin{align}
    M \!=\! F+TS \! \doteq \! \bar{M}(T,\Lambda_e) \left(1 \!+\!  \frac{\alpha \left(\! -\!  7 \Lambda_e^2 \bar{r}_h^4
 \!+\! 33 \Lambda_e \bar{r}_h^2  \!-\!  54\right)}{3 \bar{r}_h^2 \left(\Lambda_e \bar{r}_h^2+3\right)} \right)\!, \tag{B.2}
\end{align}
where $\bar{S}(T,\Lambda_e)=\frac{\pi ^2 \bar{r}_h^3}{2}$ and $\bar{M}(T,\Lambda_e)=\frac{3 \pi \bar{r}_h^2}{8}-\frac{\pi}{16} \Lambda_e \bar{r}_h^4$.

Moreover, by treating the effective cosmological constant $\Lambda_e$ and the coupling $\alpha$ as thermodynamic variables, we can write down the extended first law of thermodynamics
\begin{align}
    dM=TdS+\Theta_e d\Lambda_e + U_\alpha d\alpha, \tag{B.3}
\end{align}
where $\Theta_e$ and $U_\alpha$ denote the conjugate quantities associated with $\Lambda_e$ and $\alpha$, respectively. These conjugate quantities can be calculated utilizing the relation $dF=-SdT+\Theta_e d\Lambda_e + U_\alpha d\alpha$. Specifically, we have
\begin{align}
    \Theta_e=\frac{\partial F}{\partial \Lambda_e}\doteq -\frac{\pi  \bar{r}_h^4}{16}\left(1 \!-\! \frac{\alpha  \left(14 \Lambda_e^2 \bar{r}_h^4  \!-\!  162 \Lambda_e \bar{r}_h^2  \!+\!  360\right)}{9 \bar{r}_h^2 \left(\Lambda_e \bar{r}_h^2+3\right)} \right), \tag{B.4}
\end{align}
and
\begin{align}
   U_\alpha=\frac{\partial F}{\partial \alpha}\doteq -\frac{ \pi }{144} \left(7 \Lambda_e^2 \bar{r}_h^4-120 \Lambda_e \bar{r}_h^2+324\right). \tag{B.5}
\end{align}

Amazingly, the Wald formula \cite{Wald:1993nt,Iyer:1994ys} provides an independent approach to derive the black hole entropy. According to the Wald formula, the contribution from the Einstein--Hilbert term is $\frac{A}{4}$, where $A=2 \pi^2 r_h^3$ in the present example. And at $\mathcal{O}(\alpha)$, the contribution from $\alpha \mathcal{L}_{\text{hd}}$ can be evaluated using the unperturbed metric. This leads to
\begin{align} \label{B6}
S_W \doteq \frac{\pi^2 r_h^3}{2} + \alpha (6 \pi^2 r_h - \frac{2}{3} \pi^2 \Lambda_e r_h^3 ), \tag{B.6}
\end{align}
where $r_h$ represents the horizon radius of the perturbed black hole solution. By equating eqs.\eqref{B1} and \eqref{B6}, we obtain the relation between $\bar{r}_h$ and $r_h$ as
\begin{align}
\bar{r}_h \doteq r_h \left(1+\alpha \frac{(\Lambda_e r_h^2 - 6)^2}{3r_h^2 (\Lambda_e r_h^2 + 3)} \right). \tag{B.7}
\end{align}
Thus, by employing the relations between $\Lambda_e$ and $\Lambda$, as well as $\bar{r}_h$ and $r_h$, the expressions for $T$, $M$, $S$, $\Theta_e$, and $U_\alpha$ can also be re-expressed as the functions of another set of variables $(r_h,\Lambda, \alpha)$.   The results coincide with those obtained from traditional method \cite{Dutta:2022wbh}.


\baselineskip=1.6pt

\begin{thebibliography}{10}



\bibitem{Boulware:1985wk}
D.~G.~Boulware and S.~Deser,
\emph{String Generated Gravity Models,}
Phys. Rev. Lett. \textbf{55}, 2656 (1985).

\bibitem{Cardoso:2018ptl}
V.~Cardoso, M.~Kimura, A.~Maselli and L.~Senatore,
\emph{Black Holes in an Effective Field Theory Extension of General Relativity,}
Phys. Rev. Lett. \textbf{121}, no.25, 251105 (2018)
[arXiv:1808.08962 [gr-qc]].

\bibitem{Clifton:2011jh}
T.~Clifton, P.~G.~Ferreira, A.~Padilla and C.~Skordis,
\emph{Modified Gravity and Cosmology,}
Phys. Rept. \textbf{513}, 1-189 (2012)
[arXiv:1106.2476 [astro-ph.CO]].


\bibitem{Gubser:1998nz}
S.~S.~Gubser, I.~R.~Klebanov and A.~A.~Tseytlin,
\emph{Coupling constant dependence in the thermodynamics of N=4 supersymmetric Yang-Mills theory,}
Nucl. Phys. B \textbf{534}, 202-222 (1998)
[arXiv:hep-th/9805156 [hep-th]].

\bibitem{Caldarelli:1999ar}
M.~M.~Caldarelli and D.~Klemm,
\emph{M theory and stringy corrections to Anti-de Sitter black holes and conformal field theories,}
Nucl. Phys. B \textbf{555}, 157-182 (1999)
[arXiv:hep-th/9903078 [hep-th]].


\bibitem{Landsteiner:1999gb}
K.~Landsteiner,
\emph{String corrections to the Hawking-Page phase transition,}
Mod. Phys. Lett. A \textbf{14}, 379-386 (1999)
[arXiv:hep-th/9901143 [hep-th]].



\bibitem{Reall:2019sah}
H.~S.~Reall and J.~E.~Santos,
\emph{Higher derivative corrections to Kerr black hole thermodynamics,}
JHEP \textbf{04}, 021 (2019)
[arXiv:1901.11535 [hep-th]].

\bibitem{Ma:2023qqj}
L.~Ma, Y.~Pang and H.~Lu,
\emph{Higher derivative contributions to black hole thermodynamics at NNLO,}
JHEP \textbf{06}, 087 (2023)
[arXiv:2304.08527 [hep-th]].

\bibitem{Xiao:2022auy}
Y.~Xiao,
\emph{First order corrections to the black hole thermodynamics in higher curvature theories of gravity,}
Phys. Rev. D \textbf{106}, no.6, 064041 (2022)
[arXiv:2207.00967 [gr-qc]].

\bibitem{Cano:2019ore}
P.~A.~Cano and A.~Ruip\'erez,
\emph{Leading higher-derivative corrections to Kerr geometry,}
JHEP \textbf{05}, 189 (2019)
[erratum: JHEP \textbf{03}, 187 (2020)]
[arXiv:1901.01315 [gr-qc]].


\bibitem{Horowitz:2023xyl}
G.~T.~Horowitz, M.~Kolanowski, G.~N.~Remmen and J.~E.~Santos,
\emph{Extremal Kerr Black Holes as Amplifiers of New Physics,}
Phys. Rev. Lett. \textbf{131}, no.9, 091402 (2023)
[arXiv:2303.07358 [hep-th]].

\bibitem{Hu:2023gru}
P.~J.~Hu, L.~Ma, H.~Lu and Y.~Pang,
\emph{Improved Reall-Santos method for AdS black holes in general higher derivative gravities,} Sci. China Phys. Mech. Astron. \textbf{67}, no.8, 280412 (2024)
[arXiv:2312.11610 [hep-th]].


\bibitem{hawkingpage}
S.~W.~Hawking and D.~N.~Page, \emph{Thermodynamics Of Black Holes In Anti-De Sitter Space,} Commun. Math. Phys. 87, 577 (1983).

\bibitem{Dutta:2006vs}
S.~Dutta and R.~Gopakumar,
\emph{On Euclidean and Noetherian entropies in AdS space,}
Phys. Rev. D \textbf{74}, 044007 (2006)
[arXiv:hep-th/0604070 [hep-th]].

\bibitem{Gibbons:2004ai}
G.~W.~Gibbons, M.~J.~Perry and C.~N.~Pope,
\emph{The First law of thermodynamics for Kerr-anti-de Sitter black holes,}
Class. Quant. Grav. \textbf{22}, 1503-1526 (2005)
[arXiv:hep-th/0408217 [hep-th]].


\bibitem{Oliva:2010zd}
J.~Oliva and S.~Ray,
\emph{Classification of Six Derivative Lagrangians of Gravity and Static Spherically Symmetric Solutions,}
Phys. Rev. D \textbf{82}, 124030 (2010)
[arXiv:1004.0737 [gr-qc]].


\bibitem{Kastor:2009wy}
D.~Kastor, S.~Ray and J.~Traschen,
\emph{Enthalpy and the Mechanics of AdS Black Holes,}
Class. Quant. Grav. \textbf{26}, 195011 (2009)
[arXiv:0904.2765 [hep-th]].

\bibitem{Dutta:2022wbh}
S.~Dutta and G.~S.~Punia,
\emph{String theory corrections to holographic black hole chemistry,}
Phys. Rev. D \textbf{106}, no.2, 026003 (2022)
[arXiv:2205.15593 [hep-th]].

\bibitem{Ahmed:2023snm}
M.~B.~Ahmed, W.~Cong, D.~Kubiz\v{n}\'ak, R.~B.~Mann and M.~R.~Visser,
\emph{Holographic Dual of Extended Black Hole Thermodynamics,}
Phys. Rev. Lett. \textbf{130}, no.18, 181401 (2023)
[arXiv:2302.08163 [hep-th]].

\bibitem{Xiao:2023lap}
Y.~Xiao, Y.~Tian and Y.~X.~Liu,
\emph{Extended Black Hole Thermodynamics from Extended Iyer-Wald Formalism,}
Phys. Rev. Lett. \textbf{132}, no.2, 021401 (2024)
[arXiv:2308.12630 [gr-qc]].


\bibitem{Harlow:2022ich}
D.~Harlow, B.~Heidenreich, M.~Reece and T.~Rudelius,
\emph{Weak gravity conjecture,}
Rev. Mod. Phys. \textbf{95}, 035003 (2023)
[arXiv:2201.08380 [hep-th]].


\bibitem{Cremonini:2019wdk}
S.~Cremonini, C.~R.~T.~Jones, J.~T.~Liu and B.~McPeak,
\emph{Higher-Derivative Corrections to Entropy and the Weak Gravity Conjecture in Anti-de Sitter Space,}
JHEP \textbf{09}, 003 (2020)
[arXiv:1912.11161 [hep-th]].

\bibitem{Noumi:2022ybv}
T.~Noumi and H.~Satake,
\emph{Higher derivative corrections to black brane thermodynamics and the weak gravity conjecture,}
JHEP \textbf{12}, 130 (2022)
[arXiv:2210.02894 [hep-th]].


\bibitem{Goon:2019faz}
G.~Goon and R.~Penco,
\emph{Universal Relation between Corrections to Entropy and Extremality,}
Phys. Rev. Lett. \textbf{124}, no.10, 101103 (2020)
[arXiv:1909.05254 [hep-th]].

\bibitem{gsbook}
D.~J.~Griffiths and D.~F.~Schroeter, \emph{Introduction to Quantum Mechanics}, 3rd Edition, Cambridge University Press, (2018).

\bibitem{Ma:2024ynp}
L.~Ma, P.~J.~Hu, Y.~Pang and H.~Lu,
\emph{Effectiveness of Weyl gravity in probing quantum corrections to AdS black holes,}
Phys. Rev. D \textbf{110}, no.2, L021901 (2024)
[arXiv:2403.12131 [hep-th]].

\bibitem{Wald:1993nt}
R.~M.~Wald,
\emph{Black hole entropy is the Noether charge,}
Phys. Rev. D \textbf{48} (1993) no.8, 3427-3431
[arXiv:gr-qc/9307038 [gr-qc]].

\bibitem{Iyer:1994ys}
V.~Iyer and R.~M.~Wald,
\emph{Some properties of Noether charge and a proposal for dynamical black hole entropy,}
Phys. Rev. D \textbf{50}, 846-864 (1994)
[arXiv:gr-qc/9403028 [gr-qc]].



\end{thebibliography}
\end{document}